\documentclass[prl,twocolumn,superscriptaddress,amsmath,amssymb,aps, floatfix]{revtex4-1}

\usepackage{graphicx}
\usepackage[utf8]{inputenc}
\usepackage{epstopdf}
\usepackage{bbold}
\usepackage{booktabs}
\usepackage{color,times}
\usepackage{xcolor}
\usepackage[percent]{overpic}

\newcommand{\ba}{\begin{eqnarray}}
\newcommand{\ea}{\end{eqnarray}}
\newcommand{\be}{\begin{equation}}
\newcommand{\ee}{\end{equation}}
\newcommand{\bea}{\begin{eqnarray}}
\newcommand{\eea}{\end{eqnarray}}

\def\openone{\leavevmode\hbox{\small1\kern-3.3pt\normalsize1}}

\usepackage{theorem}

\theorembodyfont{\rmfamily}

\def\openone{\leavevmode\hbox{\small1\kern-3.3pt\normalsize1}}

\begin{document}
\title{Optimal control of many-body non-equilibrium quantum thermodynamics}
\author{N. Rach}
\affiliation{Institute for Complex Quantum Systems \& Center for Integrated Quantum Science and Technology (IQST), University of Ulm, Albert-Einstein-Allee 11, D-89069 Ulm, Germany}
\author{S. Montangero}
\affiliation{Institute for Complex Quantum Systems \& Center for Integrated Quantum Science and Technology (IQST), University of Ulm, Albert-Einstein-Allee 11, D-89069 Ulm, Germany}
\author{M. Paternostro}
\affiliation{Centre for Theoretical Atomic, Molecular and Optical
Physics, School of Mathematics and Physics, Queen's University,
Belfast BT7 1NN, United Kingdom}

\date{\today}
\pacs{xxx}

\begin{abstract}
We demonstrate the effectiveness of quantum optimal control techniques in harnessing irreversibility generated by non-equilibrium processes, implemented in unitarily evolving quantum many-body systems. We address the dynamics of a finite-size quantum Ising model subjected to finite-time transformations, which unavoidably generate irreversibility. We show that work can be generated through such transformation by means of optimal controlled quenches, while quenching the degree of irreversibility to very low values, thus boosting the efficiency of the process and paving the way to a fully controllable non-equilibrium thermodynamics of quantum processes.  
\end{abstract}

\maketitle

Non-equilibrium thermodynamics is an emerging area of investigation that focuses on processes and systems that are away from thermodynamic equilibrium~\cite{Lebon}, which is likely the most common situation encountered in nature. The consideration of explicitly time-dependent processes that is core to non-equilibrium thermodynamic is fully compatible with quantum mechanics, where there it is paramount to be able to predict and track the temporal evolution arising from quantum processes. Non-equilibrium quantum thermodynamics (NEQT)~\cite{Esposito,Campisirev} is currently being developed into a full-fledged framework with the scope of providing a thermodynamics-based analysis of non-equilibrium quantum processes. The insight gathered through such efforts ranges from fundamental~\cite{foundations} to technological issues related to the efficiency of quantum empowered micro and nano-machines~\cite{reviews}. 

In order to match the complexity of typical thermodynamic working media, and thus contribute factually to the development of a new generation of quantum technologies able to compete with existing classical machines, the attention on NEQT should be moved towards the management of complex systems. In particular, owing to the success encountered in the design of quantum simulators~\cite{Nori}, quantum many-body systems (QMBSs) hold the promises to provide us with both a sufficient degree of complexity and strong quantum features (from quantum criticality to multipartite quantum correlations), thus embodying the perfect working media for future quantum thermodynamic devices.  

The successful manipulation of quantum systems, which entails a remarkable challenge on its own, can be significantly boosted by the application of quantum optimal control (QOC) techniques~\cite{dalessandro}.   
QOC techniques implement extensive searches of the optimal steering protocol of a desired quantum dynamics  in closed and open systems by means of a functional minimization~\cite{brif2010,koch2016,rabitz2009,mukhe2013,lucas2013,mukhe2015,lloyd2014}.  Indeed, in the case of unconstrained optimizations, as considered here, QOC converges to the global optimum with respect to the fidelity with a desired goal state~\cite{rabitz2004}.
Recently, a new algorithm for QOC has been introduced to encompass many-body quantum dynamics~\cite{Doria,Caneva2011}, and a wide variety of different functionals in a straightforward way, ranging from entanglement measures to properties of time-of-flight images in cold atoms in optical lattice experiments~\cite{VanFrank2014,VanFrank2015,Scheuer2014,Rosi2013,Poschinger,Caneva2011e,Caneva2012}. It is then possible to apply such technique to develop quantum control strategies in the management of NEQT.  Indeed, while some initial steps towards on elementary quantum systems has been recently made~\cite{delcampo}, the case of QMBSs remains under-developed (despite the inherent relevance of the NEQT formalism for their dynamics~\cite{dorner}), and has so far been performed through special forms of counter-adiabatic quantum driving~\cite{qmbss}. 

In this paper we provide a strong proof of principle of the compatibility between the use of QOC approaches in QMBSs brought out of equilibrium and the principles of quantum thermodynamics. We address the paradigmatic case represented by the isolated quantum Ising model at finite temperature, and study the thermodynamic work performed on the system subjected to an external driving. We demonstrate that the {\it thermodynamic cost} of driving the system out of equilibrium, as characterized by aptly chosen quantifiers of irreversibility~\cite{reviews, Campisirev}, can be significantly lowered by the application of suitable time-dependent driving potentials, without affecting the ability of the working medium to perform a given amount of work. This provides significant evidence of the appropriateness of QOC for the harnessing on NEQT, making it a key ingredient in the design of {\it optimal} quantum thermo-machines that accomplish a given thermodynamic protocol in a finite time and with only a very small amount of dissipated resources.  

Such demonstration paves the way to the development of optimal NEQT in systems of increasing complexity and size, along the lines of 
previous demonstrations~\cite{VanFrank2014,VanFrank2015,Caneva2011e,Caneva2012}. This long-term goal will  
also benefit of the use of closed-loop optimization~\cite{Rosi2013} which would bypass the limitations introduced by increasingly complex numerical simulations~\cite{lloyd2014}, or by extrapolation 
of a given optimal strategy to the thermodynamical limit, as discussed in Ref.~\cite{Caneva2014}.   

\noindent
{\it The system \& the NEQT tools.-}We study the effects of QOC~\cite{dalessandro} on different quantifiers of irreversibility for the example of a one-dimensional Ising ring at finite (inverse) temperature $\beta$ with nearest neighbour interactions and immersed a transversal magnetic field (say along the $x$ direction). The corresponding Hamiltonian, in units of the inter-spin coupling energy, reads
\begin{equation}
\hat{\cal H}(f_t) = -f_t \sum\limits_{i=1}^N \hat \sigma_i^x + \sum\limits_{i=1}^{N} \hat \sigma_i^z \hat \sigma_{i+1}^z, \label{eq:Ising}
\end{equation}
where $\hat \sigma^k_i (k=x,y,z)$ is the $k$-Pauli operator of spin $i=1,\dots,N$ and $f_t$ denotes the dimensionless on-site energy of the spins. 
For the sake of definiteness, and without restricting the generality of our approach, we consider periodic boundary conditions such that $\sigma_{N+1}^k \equiv \sigma_1^k$. For this first approach we focus on a few-body system and compare the results achieved when driving it through a sudden quench of $f_t$~\cite{dorner} with the protocol achieved via QOC. Moreover, unless otherwise specified, we assume that the system is initially in a Gibbs state $\rho(0) = {e^{-\beta\hat{\cal H}(f_0)}}/{{\cal Z}(f_0)}$ of the initial Hamiltonian $\hat {\cal H}(f_0)$, where ${\cal Z}(f_0)$ is the partition function. 

The NEQT of an isolated system subjected to a finite-time protocol (mathematically described by the general time-evolution operator $\hat U_t$) can be predicted through the characteristic function of work distribution~\cite{Campisirev,Haenggi}
$G(u) = \text{Tr}[\hat U_te^{-iu\hat{\cal H}(f_0)}\rho(0)\hat U^{\dagger}_te^{iu\hat{\cal H}(f_t)}]$. 
The average work $\langle W\rangle$ that is done on/by the system as a result of the transformation (lasting for a time $T$) is proportional to the first moment of $G(u)$ as
\begin{equation}
{\partial_u} G(u) |_{u=0} = i\,\text{Tr}[\hat{\cal H}(f_T)\rho_T - \hat{\cal H}(f_0)\rho(t_0)] = i \langle W\rangle .
\end{equation}  
On the other hand, {\it equilibrium} properties such as the free energy difference between the initial state of the system and the {\it hypothetical} equilibrium state $\rho(t)=e^{-\beta\hat{\cal H}(f_t)}/{\cal Z}(f_t)$ at the same temperature that would have been reached through a quasi-static process  can be determined as 
$\Delta F = - \log[{\cal Z}(f_T)/{\cal Z}(f_0)]/\beta$.

We are interested in the amount of irreversibility that is generated by a non-quasistatic transformation. This provides a measure of dissipated resources du to the finite-time nature of the transformation being considered, and thus of the efficiency of the protocol itself. We thus consider three quantifiers of irreversibility, namely the irreversible entropy~\cite{foundations}, the inner friction~\cite{Friction}, and the volume entropy~\cite{campisi}, which have been introduced and studied (to different extents) in literature. By capturing different aspects of the irreversible nature of a non-equilibrium transformation, they provide different yet equally valuable thermodynamic information. We then compare the behaviour of these quantities when the chain is driven by a sudden quench of the on-site energies or by means of QOC, respectively. We show the clear superiority of the latter over the former, although comparable quantities of thermodynamic work are generated by the two strategies throughout the dynamics. 

\noindent
{\it Irreversible entropy.-}Originally introduced by Clausius, irreversible entropy has been recognised to be key in the evaluation of the efficiency of thermal machines~\cite{Velasco}. For an isothermal process, it is formally defined as
\begin{equation}
S_{\rm{irr}} = \beta(\langle W\rangle - \Delta F). 
\label{eq:Sirr}
\end{equation}
Intuitively, $S_{\rm{irr}}$ quantifies the amount of  energy that cannot be transformed in useful work due to the non-quasistatic nature of the transformation, and can be interpreted as the lag between the non-equilibrium state achieved through the dynamics and the corresponding hypothetical equilibrium one. The $2^{\rm{nd}}$ law of thermodynamics implies that $S_{\rm{irr}}\ge0$, the equality being achieved for quasi-static processes. This quantity has been recently studied for QMBSs~\cite{dorner,apollaro,brunelli} and linked to their properties when crossing critical points.  

\noindent
{\it Inner friction.-}While the irreversible entropy assumes explicitly an isothermal transformation, this obviously does not need to be the case in general. In fact, one could consider, as a second quantifier of irreversibility, the inner friction introduced in Ref.~\cite{Friction} and defined as the difference between the average work done throughout a process and the one that would have been done through a hypothetical adiabatic transformation $\langle W_a \rangle$. Explicitly
\begin{equation}
\begin{aligned}
W_{\rm{fric}} &= \langle W \rangle - \langle W_a \rangle = \text{Tr}[\hat{\cal H}(f_T)\rho(T) - \hat{\cal H}(f_T)\rho_a] \\
 &= \text{Tr}[\hat{\cal H}(f_T)\rho(T)]- \sum_m E_m(f_T)P_m^a \label{eq:wfric}
\end{aligned}
\end{equation}
with $\rho_a$ the density matrix reached at time $T$ through the adiabat, $E_m(f_T)$ the eigenvalues of the final Hamiltonian, and $P_m^a$ the probability of being in the $m^{th}$ eigenstate after the adiabatic transition. For an adiabatic process, the condition $P_m(f_0) = P_m^a$ must hold (with $P_m(f_0)=e^{-\beta E_m(f_0)}/{\cal Z}(f_0)$ the occupation probability of the $m^{\rm{th}}$ level in the initial state). We can thus simplify Eq.~\eqref{eq:wfric} as
\begin{equation}
W_{\rm{fric}} 
= \text{Tr}[\hat{\cal H}(f_T)\rho(T)] - \sum_m E_m(f_T)P_m(t_0). 
\label{eq:wfric_final}
\end{equation}
It can be rigorously proven that $W_{\rm{fric}}$  quantifies the amount of work that is dissipated when performing an adiabatic transformation in a finite time~\cite{plastina}. Such quantity is larger when the system is brought away from equilibrium. Strategies against inner friction have been considered based on ``shortcuts to adiabaticity" approaches, where control sequences are designed to minimise the distance between the actual evolution trajectory of the system and the hypothetical adiabat~\cite{delcampo}. Here we demonstrate that QOC is very effective in controlling the irreversibility quantified by $W_{\rm{fric}}$, without necessarily mimicking an adiabatic transformation. 
 
\noindent
{\it Quantum volume entropy.-}As a third measure of irreversibility, we discuss a quantum generalization of the well-known notion of classical volume entropy $S_{\rm{vol}} = \text{log}(\Phi)$,
where $\Phi$ is the volume of the phase space enclosed by the constant energy surface of a classical thermally isolated system. Such generalisation, which has been formulated in Ref.~\cite{campisi}, requires the definition of the instantaneous number operator
$\hat {\cal N}(t) = \sum_{k=0}^{K} k |k(t)\rangle\langle k(t)|$  
with $\{|k(t)\rangle\}$ the set of instantaneous eigenstates of a non-degenerated time-dependent Hamiltonian $\hat{H}(t)$. 
In terms of $\hat{\cal N}(t)$, the quantum volume entropy is defined as~\cite{campisi}
\begin{equation}
S_{\rm{Qvol}} = \text{Tr}[\rho(T)\hat{\cal S}(T) - \rho(0)\hat {\cal S}(0)] \label{eq:S_vol}
\end{equation}
with $\hat {\cal S}(t) = \log(\hat{\cal N}(t) + \mathbb{1}/2)$. While it can be proven that $S_{\rm{Qvol}}\ge0$, it is worth mentioning that in the spectrum of the Ising model degeneracy may arise for specific values of $f_t$. However, as Eq.~\eqref{eq:S_vol} would only depend on the initial and final value of $f_t$, we can avoid these cases by choosing suitable boundary conditions. In addition we introduce a slight change in Eq.~\eqref{eq:S_vol} to deal with the degenerate cases~\cite{nota} and consider the operator
$\hat{\cal N}_d(t) = \sum_{k=0}^{\tilde{K}} k \sum_{l=1}^L |k_l(t)\rangle\langle k_l(t)|$, 
where $\tilde{K}$ is the number of non-degenerated eigenvalues and $L$ is the order of degeneracy. We thus simply weight states with the same eigenvalue with the same $k$. For an adiabatic transformation $S_{\rm{Qvol}}=0$, thus sharing similarities with $W_{\rm{fdic}}$ (cf. the results of our numerical investigation).

 \begin{figure}[t]
\centerline{{\bf (a)}\hskip4.0cm{\bf (b)}}
\begin{overpic}[width=0.250\textwidth]{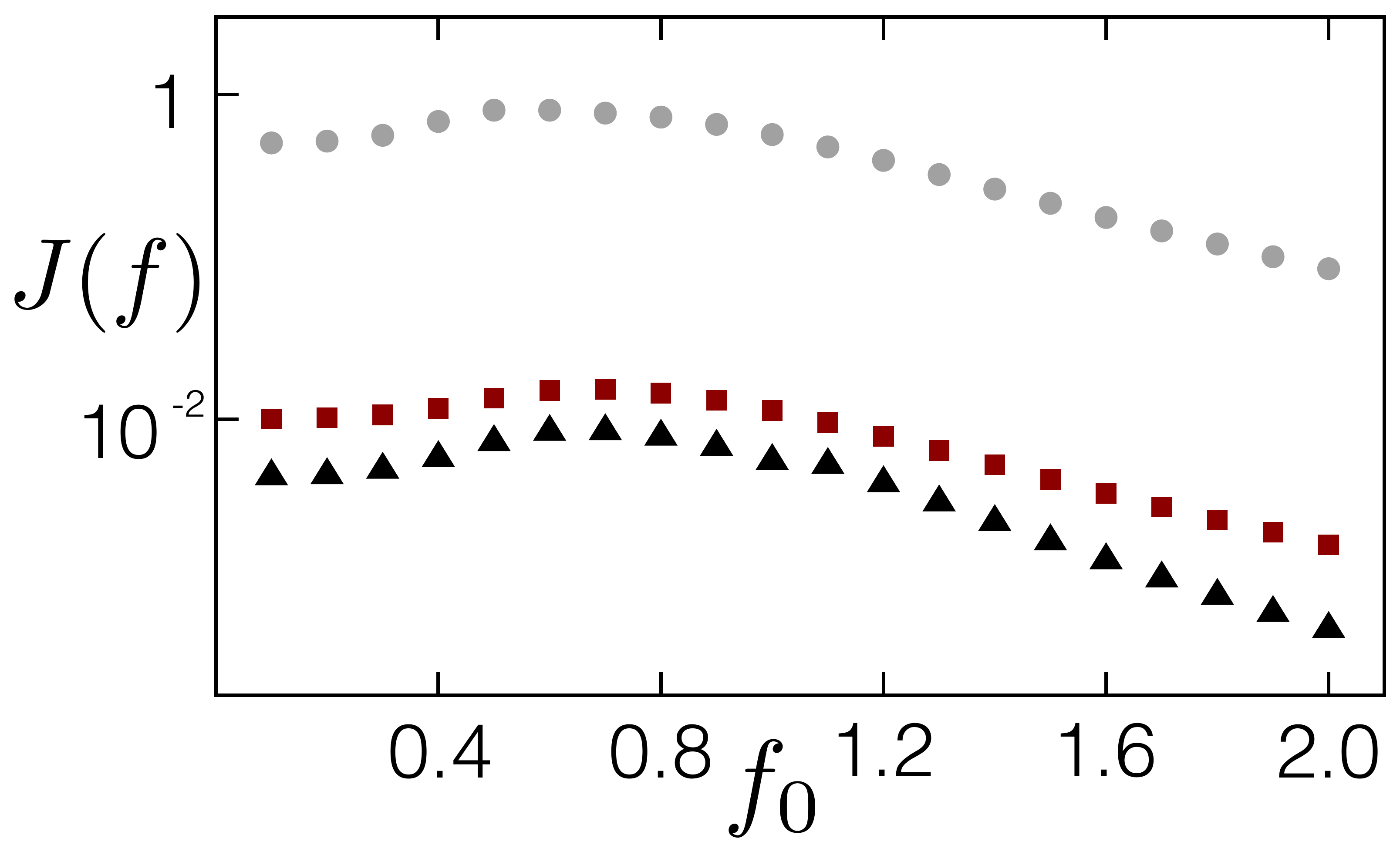}
\end{overpic}\begin{overpic}[width=0.237\textwidth]{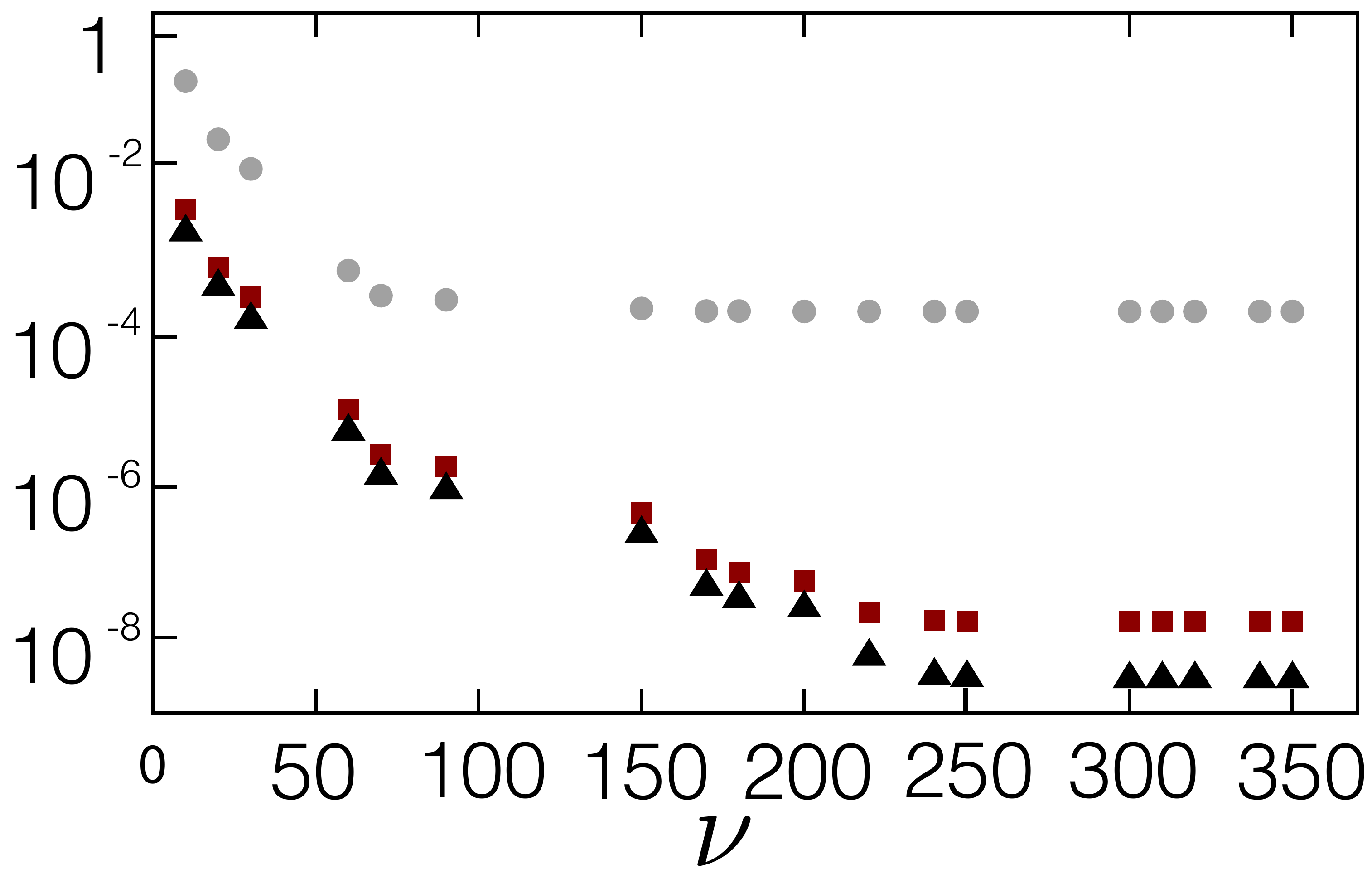}
\end{overpic}
\caption{(Color online) {\bf (a)} Plot of $S_{\rm{irr}}$ (grey dots), $W_{\rm{fric}}$ (red squares) and $S_{\rm{Qvol}}$ (black triangles) against the initial control $f_0$ for a sudden quench with $N=4$ and $\beta=50$. The kink in the curve of the volume entropy at $f_0=1$ is due to the crossing of the energy levels at this specific point. As this case is excluded in Eq.~\eqref{eq:S_vol}, we use the definition of $\hat{\cal S}(t)$ in terms of $\hat{\cal N}_d(t)$. Note that although it provides physically reasonable results, the effect is still visible in the curve. {\bf (b)} Plot of $S_{\rm{irr}}$ (grey dots), $W_{\rm{fric}}$ (red squares) and $S_{\rm{Qvol}}$ (black triangles) against the number of function evaluations ($\nu$) for the optimization procedure. 
 Even when $S_{\text{irr}}$ levels to its lowest value, the other two quantifiers get minimized further. The parameters used int he simulations are $N=4$, $\beta=50$ and $f_0 = 0.8$.}\label{fig:N4sq}
\end{figure}

\noindent
{\it Study \& results.-}We investigate the amount of irreversibility generated throughout the process $\hat{\cal H}(f_0) \to \hat{\cal H}(f_T)$ for different protocols $f_t$ with $t \in [t_0,T]$, starting with a sudden quench. This protocol is based on a quick change in the control $f_t$ that is too fast for the state to follow. Thus, the time scale in which $f_t$ changes has to be much smaller than the typical time scale of the system dynamics. Throughout the simulation, the sudden quench is described by the Heaviside function
\begin{equation}
f_t = \begin{cases}
		f_0, \hspace{0.2cm} 0\le t \leq T, \\
	    f_T, \hspace{0.2cm} t > T.
	   	\end{cases}
\end{equation}  
Fig.~\ref{fig:N4sq} {\bf(a)} reports the results for a system with $N=4$ particles, an inverse temperature $\beta = 50$ and $f_T = f_0 + 0.1$. It shows the different quantifiers of irreversibility as a function of $f_0$. Qualitatively, we observe a similar behaviour of the irreversible entropy as described in~\cite{dorner} for larger systems. In addition, we see that the inner friction and the entropy given by Eq.~\eqref{eq:S_vol} follow the curve of the irreversible entropy although the function values differ from each other which makes them similar but not equal.  
In a next step, we compare these results to the ones of an optimized control $f^*_t$ determined by use of the dCRAB method~\cite{Rach}. It expands the control $f_t$ in the $j^{\rm{th}}$ super-iteration as 
$f^j_t = g^j_t + \frac{1}{\lambda_t}\sum\limits_{k=1}^{N_c} c_k^j \sin(\omega_k^j t + \phi_k^j)$ 
with 
$g^j_t = f^{j-1}_t$ and $g^1_t = f_0 + (f_T-f_0)t/{T}$. 
The coefficients $c^j_k$ are optimized with respect to a figure of merit throughout each super-iteration by the Nelder-Mead simplex method~\cite{NelderMead} for random frequencies $\omega_k^j \in [0,\omega_{max}]$. The phase $\phi^j_k \in \{0, \frac{\pi}{2}\}$ is also chosen randomly to switch between sine and cosine, and the function $\lambda(t)$ is used to impose the boundary conditions at $t=0$ and $T$. The optimization is considered complete as soon as one of the following criteria is fulfilled: 
{\it 1)} The value of the cost function is below an error threshold $\eta_e$;
{\it 2)} A certain number of maximal super-iterations (frequencies) is reached;
{\it 3)} The relative change within one super-iteration $1 - J(f^j)/J(f^{j-1})$ is below a change threshold $\eta_c$.
Unless otherwise mentioned, we set $\eta_{e,c}= \eta$, $T\omega_{max}/2\pi = 20$ and $J(f) = S_{\rm{irr}}$ here. In addition we use $N_c = 4$ frequencies in each super-iteration and fix the final (numerical) time to $T=\pi$. 
Fig.~\ref{fig:N4sq} {\bf (b)} reports the results of a simulation with $N=4$ particles, $\eta = 10^{-5}$, $f_0=0.8$ and $f_T = f_0 + 0.1$. It shows the values of all three quantifiers against the number of function evaluations $\nu$. The performance of the optimization is quite informative and inspires a set of interesting observations. First, we notice that, despite being designed upon the irreversible entropy $S_\text{irr}$, the optimization protocol is able to reduce the values taken by all the three irreversibility quantifiers that we have considered. Therefore, from the standpoint of QOC, the various figures of merit used to characterize thermodynamics irreversibility are broadly equivalent. Second, even in the range where the $S_{\text{irr}}$ reaches its lowest value, the other two quantifiers still show a significant decrease. This highlights the inherent differences of the three figures of merit here at hand, which are captured by the control protocol that we have identified. Remarkably, the results gather through our analysis seem to suggest that, although not designed to mimic a transitionless dynamics, the application of QOC to this problem results in an approximately adiabatic effective evolution. In fact, both $W_{\text{fric}}$ and $S_{\text{Qvol}}$ are nullified along reversible adiabats, while $S_{\text{irr}}$ is not, in general. This leaves room for interesting quantitative comparisons with the performance of transitionless quantum driving protocols~\cite{delcampo,qmbss}.

\begin{figure}[t]
{\bf(a)}\hskip4.5cm{\bf (b)}
\begin{overpic}[width=0.24\textwidth]{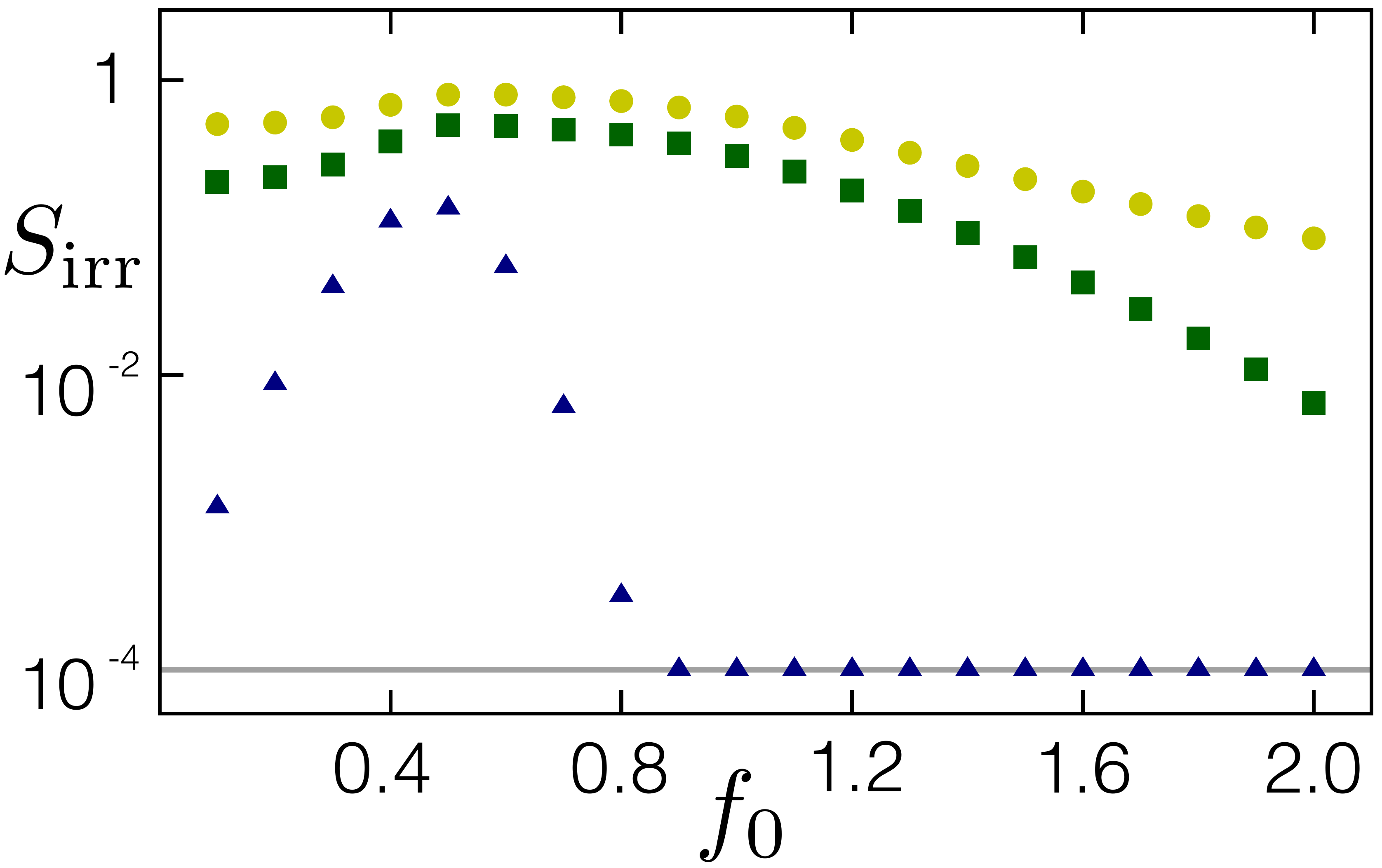}
\end{overpic}~\begin{overpic}[width=.24\textwidth]{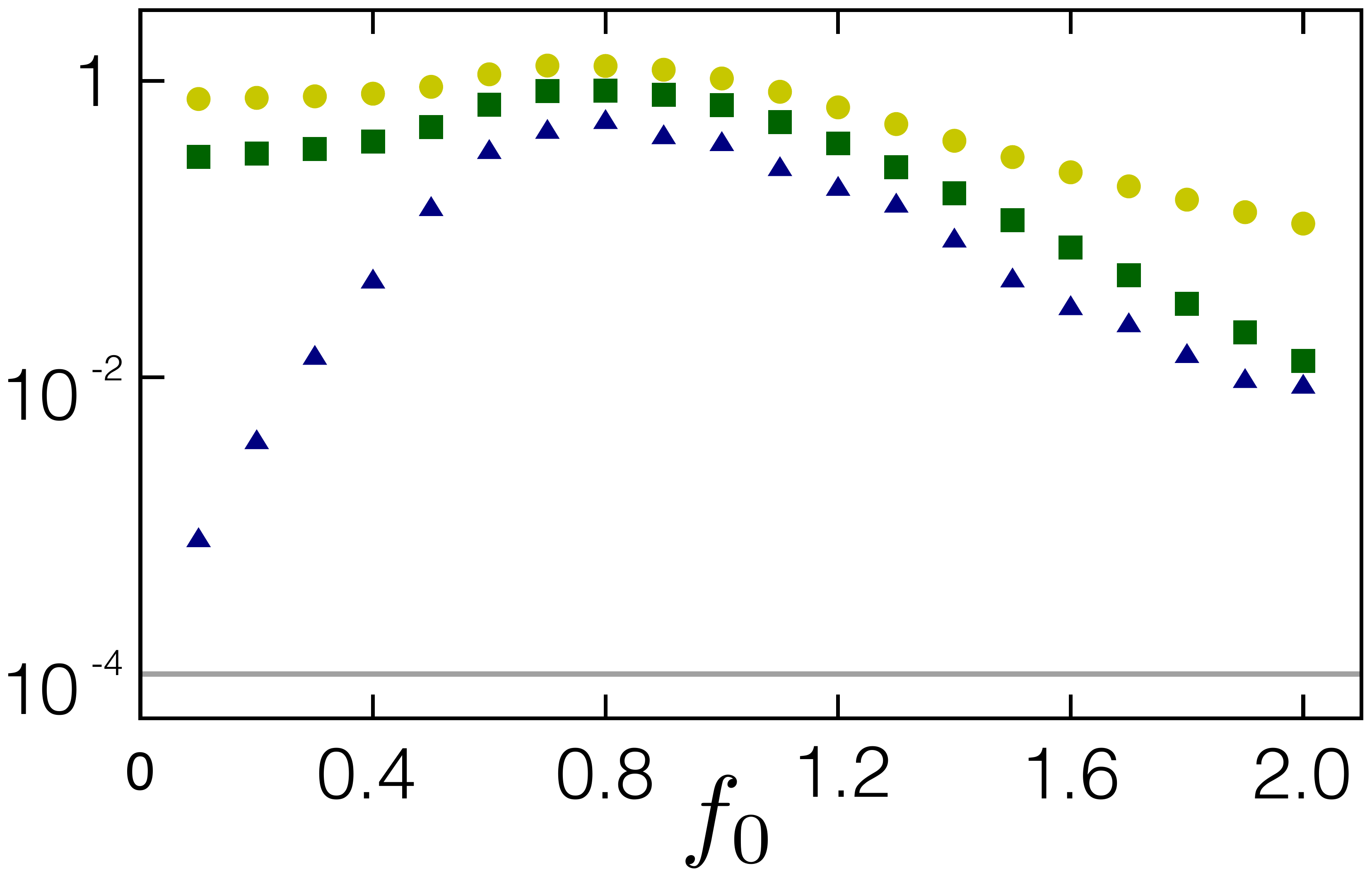}\end{overpic}
\caption{(Color online) {\bf (a)} Plot of $S_{\rm{irr}}$ as a function of the initial control $f_0$ of a sudden quench (yellow dots), a linear ramp (green squares) and an optimized control $f^*_t$ (blue triangles) for $N=4$ and $\beta=50$. The grey line indicates the error threshold $\eta=10^{-4}$. {\bf (b)} The use of the optimal driving protocol designed for panel {\bf (a)} applied to a chain of $N=6$ spins in the same operating conditions. \label{fig:N4sq_vs_qoc}}
\end{figure}

In a next step we compare the results for $S_{\text{irr}}$ following the sudden quench to those corresponding to the linear ramp and to an optimized pulse $f^*_t$. For this simulation we decrease the final time to $T=\pi/4$ and look at systems with $N=4$ particles. Fig.~\ref{fig:N4sq_vs_qoc} {\bf (a)} shows $S_{\rm{irr}}$ as a function of $f_0$ for the sudden quench (yellow dots), the linear ramp (green squares) and the optimal control (blue triangles). Clearly, although the linear ramp is slightly better than the sudden quench, the optimized pulse leads to an improvement, which can be rather substantial and resulting in a reduction of the irreversible entropy of up to three orders of magnitude with respect to the quenched case. Moreover, we have investigated the reduction of irreversibility when  the optimal protocol designed for a small system is applied, with no changes, to a bigger one. We have found that a reduction of $S_{\text{irr}}$ is indeed still possible, although this was not obvious {\it a priori}, thus demonstrating the relative robustness of the driving protocol to changes in the system's size. In particular, Fig.~\ref{fig:N4sq_vs_qoc} {\bf (b)} shows the results achieved by applying the optimal protocol designed for $N=4$ to a $6$-spin system. While the performance for $f_0\gtrsim0.8$ is very close to that of a linear ramp, smaller values of $f_0$ are associated with reductions of $S_{\text{irr}}$ comparable to the optimal case in Fig.~\ref{fig:N4sq_vs_qoc} {\bf (a)}.

\begin{figure}[b]
 {\bf (a)}\hskip4.5cm{\bf (b)}
\begin{overpic}[width=0.4\textwidth]{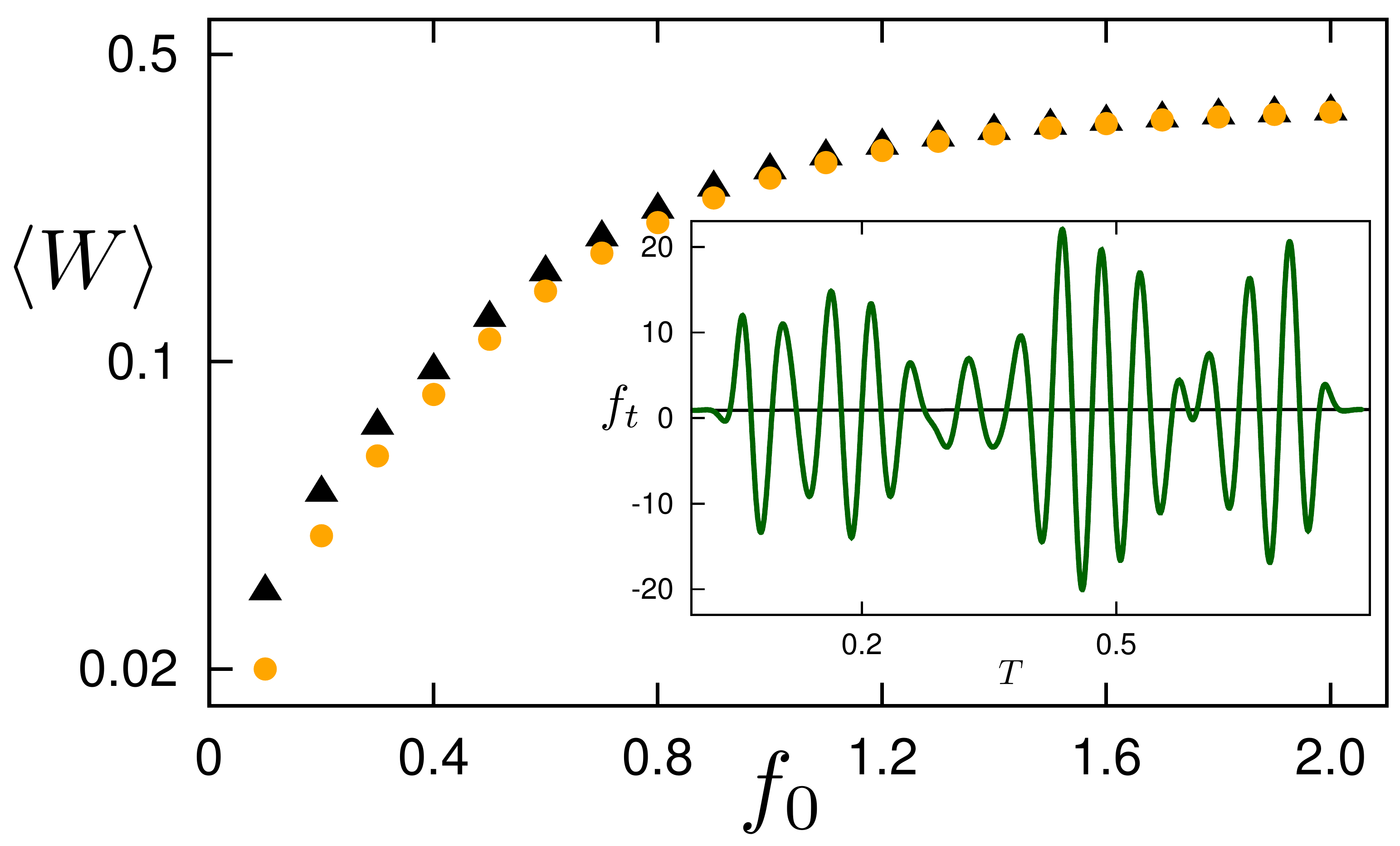}
\end{overpic}
\caption{(Color online) {\bf (a)} Plot of $\langle W\rangle$ as a function of $f_0$ for the sudden quench (orange dots) and the optimized pulse (black triangles). We see that the amount of work performed on the system is in both cases in the same range. Inset: Plot of the optimal control $f^*_t$ (green line) and the initial guess function $g^1_t$ (black line). The optimal control was achieved with three super-iterations (12 frequencies) of the dCRAB method for a system of $N=4$ spins at $\beta=50$. The correspondent (optimized) value of the cost function is (numerically) zero which means below the threshold $\eta$.}\label{fig:work}
\end{figure}

Moreover, we compare the amount of work done to the system throughout the process for the sudden quench and the optimal control. The main panel of Fig.~\ref{fig:work} shows  $\langle W\rangle$ against $f_0$ for the sudden quench (orange dots) and the optimal control (black triangles) in a system with $N=4$  and $\beta=50$. We see that the amount of work is in the same order of magnitude although the losses (quantified by $S_{\text{irr}}$) could be significantly reduced in the case of the optimal control. 
We emphasise that although only the plots for $N=4$ particles are shown, the discussed effects are observed for the cases of $N=3$ and $N=5$ as well which implies that the effectiveness of our strategy is not related to this specific size of the system and more complex situations can be addressed successfully. The inset of Fig.~\ref{fig:work} shows the optimized pulse for $N = 4$, $\beta = 50$ and $f_0 = 0.9$. The pulse, although more complex than a simple linear ramp, is clearly smooth and with finite bandwidth, thus in principle easily implementable in the lab. Moreover, when needed, additional constraints can be included in the optimization to match the experimental requirements. Finally, also robustness with respect to unavoidable limited knowledge of the system and/or errors in the control field can be taken into account to improve the final result.  

\noindent
{\it Conclusions.-}We have demonstrated the effectiveness of QOC approaches to NEQT problems applied to interacting few-body systems. Our study shows that substantial amount of thermodynamic work can be produced by driving a multi-particle system through a finite-time QOC-based protocol that substantially reduces the irreversibility resulting from the non quasi-static nature of general quantum transformations. Remarkably, QOC is able to quench figures of merit addressing different facets of irreversibility, thus providing unforeseen leverage for the control of NEQT~\cite{qmbss}. This could be crucial for the design of quantum cycles and thermo-machines operating at very low degrees of entropy productions, yet still able to produce significant amounts of work. Although we only addressed small-sized problems, we have strong numerical evidences that even the presented optimization scheme could be used even for larger system. In fact, computationally our problem is not different from state-to-state transfer problems in QMBSs, which have been successfully optimized~\cite{Caneva,Doria,Pichler}. 
 
\noindent
{\it Acknowledgements.-}NR is grateful to the Centre for Theoretical Atomic, Molecular, and Optical Physics, School of Mathematics and Physics, Queen's University Belfast, for hospitality during the completion of this work, and thanks Maximilian Keck for the enlightening discussions. We acknowledge support from the EU Projects
TherMiQ, and RYSQ, the John Templeton Foundation (grant number 43467), the Julian Schwinger Foundation (grant number JSF-14-7-0000), the UK EPSRC (grants number EP/K029371/1 and EP/M003019/1), and the DFG via the SFB/TRR21. This work was partially performed on the computational resource
bwUniCluster funded by the Ministry of Science, Research and Arts and the Universities of the State of Baden-W\"urttemberg (Germany) within the framework program bwHPC. We acknowledge partial support from COST Action MP1209.

\end{document}